\documentstyle[11pt,fleqn]{article}
\def\ref{par\noindent\hangindent=6mm\hangafter=1}
\begin{document}
\baselineskip 8mm

\begin{center}
{\bf COSMOLOGICAL LEVINSON THEOREM}

\bigskip

 H. Rosu\footnote{E-mail: rosu@ifug3.ugto.mx}    

{\it Instituto de F\'{\i}sica, Universidad de Guanajuato,
Apdo Postal E-143, Le\'on, Gto, M\'exico}

\end{center}

\bigskip

{\bf Summary.} - If at least some Wheeler-DeWitt solutions can be interpreted
as zero-energy resonances then the total s-wave cross section of the
corresponding quantum universes is infinite.

{\scriptsize PACS 04.60 - Quantum gravity
\hspace{1.5cm} gr-qc/9712072
\hspace{1.5cm} Nuovo Cimento B 114, 113-114 (Jan. 1999)}




The present, common `quantum'
cosmological framework is based on the Wheeler-DeWitt equation, which being
a `stationary', zero-energy differential equation appears to have no obvious
time dependence and therefore forcing one to invent clock models, a feature
known in the literature as the problem of time.

Here I would like to draw attention to the fact that for a more advantageous
study of some important issues, like quantum cosmological irreversibility,
it may prove convenient to think of some Wheeler-DeWitt
`wavefunction of the universe' solutions
as similar to zero-energy resonances (also known as half-bound states,
and possible only for zero angular momentum \cite{nm}) in ordinary quantum
scattering.
The zero-energy resonance wavefunctions are zero at the arbitrary chosen
cosmological origin, are finite at infinity, and are not normalizable.
For such a zero-energy resonance, taking into account that there are no
cosmological `bound states', one may write a cosmological nonperturbative
Levinson theorem of the form \cite{nm}
$$
\eta _{u}(0)=\frac{\pi}{2}\sin ^{2}\eta _{u}(0)
\eqno(1)
$$
for the $s$-wave phase shift of the quantum universe
at zero energy, with the  obvious
result $\eta _{u}(0)=\frac{\pi}{2}$. Thus, one can claim that
the total s-wave scattering cross section of the `quantum' universes with
zero-energy resonances is infinite.

\bigskip
{\bf Acknowledgment}. -
Work partially supported by the CONACyT Project 458100-25844E.



 \end{document}